# Phase modulation parallel optical delay detector for microwave angle-of-arrival measurement with accuracy monitored

Z. Cao,[1,*] Q. Wang,[1] R. Lu,[1,2] H.P.A. van den Boom,[1] E. Tangdiongga,[1] and A.M.J. Koonen [1]

[1] COBRA Institute, Eindhoven University of Technology, NL 5600 MB Eindhoven, The Netherlands
[2] School of Optoelectronic Information, State Key Laboratory of Electronic Thin Films and Integrated Devices, University of Electronic Science and Technology of China, Chengdu 610054, China
*Corresponding author: z.cao@tue.nl

A novel phase modulation parallel optical delay detector is proposed for microwave angle-of-arrival (AOA) measurement with accuracy monitored by using only one dual-electrode Mach-Zehnder modulator. A theoretical model is built up to analyze the proposed system including measurement accuracy monitoring. The spatial delay measurement is translated into the phase shift between two replicas of a microwave signal. Thanks to the accuracy monitoring, the phase shifts from 5° to 165° are measured with less than 3.1° measurement error.
http://dx.doi.org/10.1364/OL.99.099999

Determining the location of a microwave signal is of great importance for retrieving the position of objects. The parameter angle-of-arrival (AOA) or equivalently the time difference of arrival (TDOA) is required to accurately identify the position. An optical approach to measure the AOA can offer many benefits due to its intrinsic features like ultra-low loss and huge bandwidth, which allows high accuracy, and immunity to electromagnetic interferences. Moreover, with the rapid development of ultra-low drive voltage electro-optical modulators (EOMs) [1,2] and high-speed photo-diodes [3,4], barriers between electrical domain and optical domain are gradually eliminated. Recently some photonic approaches are proposed to measure AOAs of microwave signals [5-8]. Some of these approaches are based on optical modulators with the advantage of the availability of mature and commercial products. Furthermore, such kind of schemes are scalable based on integrated optics. In Ref.[8], a serial optical delay detector using two EOMs and one discrete optical delay line is proposed for AOA measurements. The discrete optical delay line (fiber) between EOMs will introduce unwanted interferences from environment variations (e.g. temperature). In Ref. [9], a parallel optical delay detector (PODD) is proposed based on a parallel Mach-Zehnder modulator (P-MZM). Its integrated structure can increase the tolerance toward environment variations. Since there are three DC-biases and only one monitored parameter (power of optical carrier) in a P-MZM, the simple and robust automatic bias control (ABC) is difficult to achieve. To solve this problem, the core idea is to avoid unnecessary DC-biases by replacing intensity modulation with phase modulation. In this letter, a novel phase modulation parallel optical delay detector (PM-PODD) using only one dual-electrode Mach-Zehnder (DE-MZM) is proposed. Because there is only one DC-bias in a DE-MZM, the simple and robust ABC is achievable. Moreover, the complexity and intrinsic insertion loss of the proposed scheme are halved compared to the one in Ref. [9].

The proposed scheme for AOA (or TDOA) measurement is depicted in Fig.1. It includes a DFB laser, a DE-MZM, an optical notch filter (ONF), and two optical power meters. The upper arm (U-arm) and lower arm (L-arm) are the two arms inside DE-MZM. Bias-1 is the tunable phase shifter between U- and L-arm. The U- and L-arm of the DE-MZM are connected to two antennas, Ante-1 and Ante-2. The distance between Ante-1 and Ante-2 is denoted as $d$. The AOA is denoted as $\psi$ and the corresponding TDOA can be expressed by:

$$t = d\cos(\psi)/c \quad (1)$$

where $c$ is the light velocity in air. Properties of the electrical path between Ante-1 and U-arm can be different in length and in impedance to the one between Ante-2 and L-arm. This difference can introduce fixed phase offsets for different frequencies. Such phase offsets can be easily compensated using a look-up table. The TDOA $\tau$ will introduce the phase shift $\varphi$ between Ante-1 and Ante-2 as shown:

$$\varphi = \tau \times \omega_m \quad (2)$$

where $\omega_m$ is the angular frequency of microwave signal. Therefore, the task of the proposed PM-PODD scheme is to measure the phase shift $\varphi$ in optical domain. The phase shift $\varphi$ caused by spatial delay $\tau$ will be translated to the phase difference of optical sidebands. The following task is to measure this phase difference of optical sidebands by using optical power meters. The U-arm is biased at the null point to suppress the optical carrier. The lightwave from the CW laser is modulated by two replicas of the microwave signal with phase shift $\varphi$ at the U- and L-arm, of which the spectra are shown in Fig. 1(a) and (b), respectively. The output optical signals from both U- and L-arm with phase shift $\varphi$ are then combined with an additional phase shift $\theta$ induced by the bias voltage applied to Bias-1. The optical spectrum of the combined signal is shown in Fig.1 (c). As shown in Fig. 1(d), the optical carrier is separated from the optical sidebands via an optical notch filter (ONF-1 shown in Fig. 1). The spectrum of filtered optical sidebands is shown in Fig. 1(e).

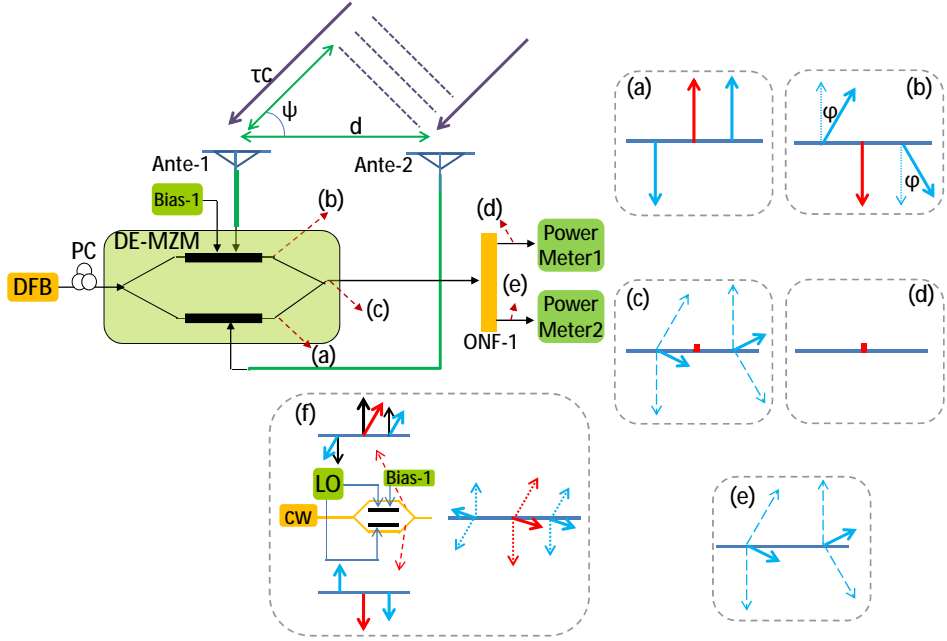

Fig. 1. The principle of AOA measurement based on parallel optical delay detector.

Now we deduce a theoretical model for the output optical power regarding the phase shift $\varphi$. The optical carrier can be expressed as:

$$E(t) = E_0 \exp(jw_0 t) \qquad (3)$$

where $E_0$ and $\omega_0$ are the amplitude of the electrical field and the angular frequency of optical carrier, respectively. The optical carrier is then split into two arms (U- and L-arm). The microwave signal applied to the U- and L-arm can be expressed as:

$$\begin{aligned} E_{am}(t) &= E_m \exp(jw_m t) \\ E_{bm}(t) &= E_m \exp(jw_m t + j\varphi) \end{aligned} \qquad (4)$$

The optical signal after the DE-MZM can be expressed as:

$$E_{out}(t) = \frac{1}{2} E_0 \exp(jw_0 t) \times \sum_{n=-\infty}^{\infty} [\exp(jn\varphi) + \exp(j\theta)] j^n J_n(m) \exp(jnw_m t) \qquad (5)$$

where $m = \pi E_m / V_\pi$ denotes the modulation depth. $\theta$ is a phase shift introduced by a DC bias. This $\theta$ is equal to $\pi$ for the null points. The high order (>2nd) sidebands are ignored since the received microwave power of these high order sidebands is relatively low. The expression can be further written as:

$$\begin{aligned} E_{out}(t) = &+\frac{1}{2} j E_0 J_{+1}(m)[\exp(j\varphi)-1]\exp(jw_0 t + jw_m t) \\ &-\frac{1}{2} j E_0 J_{-1}(m)[\exp(-j\varphi)-1]\exp(jw_0 t - jw_m t) \end{aligned} \qquad (6)$$

After the optical notch filter, the power of the upper sideband can be obtained as:

$$\begin{aligned} P_{+1} &= \frac{1}{4} E_0^2 J_{+1}^2(m)[\exp(j\varphi)-1][\exp(-j\varphi)-1] \\ &= \frac{1}{2} E_0^2 J_{+1}^2(m)[1-\cos(\varphi)] \end{aligned} \qquad (7)$$

where $J_{+1}(m)$ is the Bessel function of first kind with regard to modulation index (m). Similarly, we can obtain the power of lower sideband as:

$$\begin{aligned} P_{-1} &= \frac{1}{4} E_0^2 J_{-1}^2(m)[\exp(-j\varphi)-1][\exp(j\varphi)-1] \\ &= \frac{1}{2} E_0^2 J_{-1}^2(m)[1-\cos(\varphi)] \end{aligned} \qquad (8)$$

Thus, the power of the upper and lower sideband can be written in a unified expression:

$$P_{\pm 1} = \frac{1}{2} E_0^2 J_{\pm 1}^2(m)[1-\cos(\varphi)] \qquad (9)$$

It is clear that the output power is related to the phase shift $\varphi$. Since $J_{+1}^2(m)$ is equal to $J_{-1}^2(m)$, the output power of upper and lower sidebands induced by phase shift $\varphi$ are equal. This feature will be employed for the measurement of the two samples (output power samples of both the upper and lower sidebands) with high robustness since the noise is averaged. The upper and lower sidebands do not need to be separated, therefore an optical notch filter can be used to obtain the wanted results. From Eq. 9, it also indicates that the amplitudes of the sidebands are related to the modulation index m. The high order sidebands are negligible for low driving power, which is the case for AOA measurements. The value required for the AOA estimation is the normalized power ($P_n$), thus the value of $E_0$ and $J_{\pm 1}(m)$ are less interesting. We can obtain the expressions for the TODA ($\tau$) and AOA ($\psi$) as:

$$P_n = P_m / P_0, \quad j = \arccos(P_n - 1)$$
$$t = \arccos(P_n - 1)/w_m, \quad y = \arccos(tc/d) \quad (10)$$

According to Eq.10, to estimate the values of $\tau$ and $\psi$, the required parameters are $P_n$ and $\omega_m$. $P_n$ can be obtained by measuring $P_m$ and $P_0$. $P_0$ is the measured output power with zero phase shift ($\varphi=0$) and the calibration procedure will be detailed in the following. Based on the measured $P_m$, the phase shift $\varphi$ can be estimated for a given value of $\omega_m$. Further we can get the AOA (or TDOA) based on Eq. 10. If $\omega_m$ is unknown, an additional photonic scheme can be utilized to perform a frequency measurement before the AOA (or TODA) measurement. In the above discussion, we assume that the optical carrier is well suppressed, and thus the power and phase shift can be fully modeled according to Eq. 10. However, both the limited extinction ratio and the DC drift will introduce measurement errors. Since the limited extinction ratio is given once the modulator is fabricated, we emphasize on the analysis of DC drifts induced measurement errors. As shown in Fig. 1(f), the proper bias applied to the U-arm for optical carrier suppression should introduce $\pi$ phase shift between the U- and L-arm as the black arrows shown in Fig. 1(f). The DC drift will introduce the phase shift (θ) to the optical carrier and the sidebands. The E-field of output optical signal can be expressed as:

$$E_{out}(t) = \frac{1}{2} E_0 \exp(jw_0 t)[1+\exp(jq)]J_0(m)$$
$$+ \frac{1}{2} E_0 J_{+1}(m) \exp(jj + jq) \exp(jw_0 t + jw_m t) \quad (11)$$
$$+ \frac{1}{2} E_0 J_{-1}(m) \exp(-jj + jq) \exp(jw_0 t - jw_m t)$$

Comparing with Eq.6, the power of the sidebands is not accurate to present the phase shift $\varphi$ with such unwanted $\theta$. Both DC drifts in the U- and L-arm will introduce similar effects. Thus it is of interest to monitor the DC drift during the measurement. Since the DC drift simultaneously introduces residual leakage of the optical carrier, the power measurement of the optical carrier can be used to monitor the DC drift. As shown in Fig.1, an optical notch filter (ONF-1 shown in Fig.1) is employed to deeply separate the optical carrier and the sidebands. The separated optical carrier can then be monitored during the measurement process. In a practical system, an automatic bias control circuit can be used to reduce DC drifts. Such kind of scheme is widely available since they are previously used for stable advanced modulation format generations.

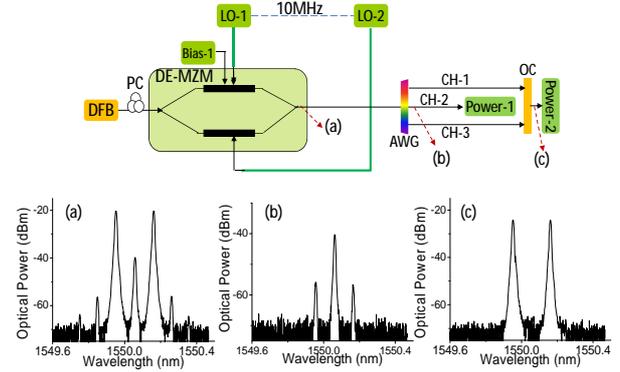

Fig. 2. The experimental setup of AOA measurement based on phase modulation parallel optical delay detector.

Fig. 2 shows the proof-of-concept experimental setup of the AOA (or TDOA) measurement based on the proposed PM-PODD. The optical carrier is generated from a DFB laser at 1550.016nm with 1dBm power. It is fed into a DE-MZM after a polarization controller (PC). Two commercial microwave sources (LO-1 and LO-2) are employed to drive the U- and L-arm at a frequency of 12.5GHz, respectively. A 10MHz sinusoidal signal generated from LO-1 is sent to LO-2 for phase synchronization. DE-MZM is biased at the null of its power transfer curve and the optical spectrum of the combined signal is shown in Fig. 2(a). The 2nd and 3rd order sidebands are observed > 35-dB and > 45-dB lower than the 1st sidebands, respectively. Thus the higher order sidebands can be neglected. The phase differences between LO-1 and LO-2 induced by different electrical paths and impedance mismatches are measured by a sampling oscilloscope (digital communication analyzer). It is then further calibrated via a look-up table. The optical output signal is then separated by an array waveguide grating (AWG) which acts as an optical notch filter. In general, any kind of optical notch filters can be used here if its stop-band is narrower than the carrier frequency of measured microwave signal. The channel spacing of the AWG is 12.5GHz and the signal is then separated into three channels. The optical carrier is in the middle channel (noted as CH-2) and two sidebands are in two neighbor channels (noted as CH-1,3). The signal from CH-2 is used for the DC drift monitoring (measured at Power-1) with its spectrum shown in Fig. 2(b). Optical signals from CH-1,3 are then coupled again via a 3-dB coupler (OC) for power measurement (at Power-2). Its spectrum is shown in Fig. 2(c). The measured spectra from CH-1,3 with different phase shifts are shown in Fig. 3. As shown in Fig. 3(a), the sidebands are enhanced when the phase shift is 180°. The sidebands are destructively combined once the phase shift is 0° shown in Fig. 3(c). We can clearly observe that the power of sidebands degrades when the phase shift decreases, which coincides with Eq.9. It should be noted that the 2nd order sidebands vary when the phase shift changes. In our experiment, the 2nd order sidebands are filtered out as well by the AWG as shown in Fig. 2(c). Thus its influence will be largely limited. However, even for the system using only one optical notch filter, its effect is negligible since their power is >35dB lower than the power of 1st order sidebands. A calibration to obtain the minimum output power of optical carrier is carried out. After that, the $P_0$ is obtained by getting the maximum power of sidebands. In Fig. 3(a)-(c), the optical carrier is not completely suppressed mainly due to the limited extinction ratio, since the DC drift is mostly eliminated at the initial calibration stage. Two measurements are carried out. The first measurement is carried out just after the initial calibration stage. The power of the filtered optical carrier ($P_0$) is -45.3dBm. We measure the phase shift φ and corresponding measurement errors. The second measurement is after a few tens of minutes and $P_0$ increased to -40.5dBm due to DC drift. The phase shift φ is measured again with measurement errors. The measured powers versus different phase shifts (circles) are shown in Fig. 4(a) for the first measurement. The theoretical power distribution (red curve) versus phase shift is also shown in Fig. 4(a). An accepted agreement is obtained. The detailed measurement errors are shown in Fig. 4(b) and (c) for both measurements. The measurement errors are less than 3.1° within the range from 5° to 165° when the filtered power is -45.3dBm. It is clear that the measurement errors increase when the phase shift goes to 0°. This is mainly because of imperfect destruction of sidebands induced by the limited extinction ratio. Comparing Fig. 4 (b) and (c), it is obvious that the measurement errors increase up to 7.7° when $P_0$ drifts to -40.5dBm. It shows that the measurement accuracy degradation induced by DC drifts can be well monitored in this scheme.

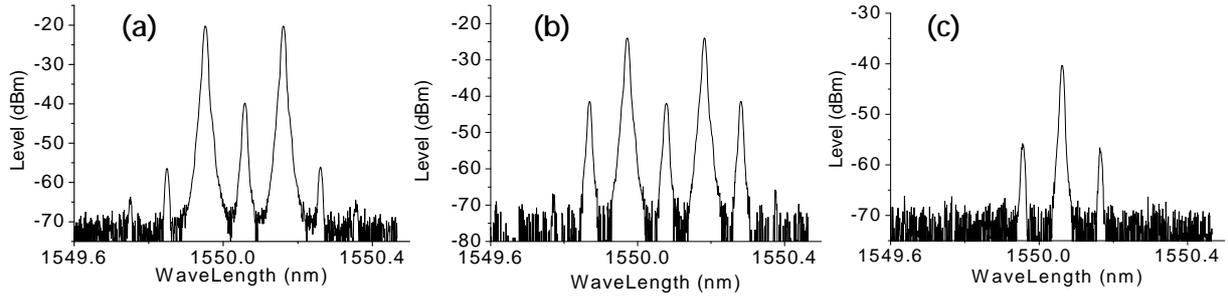

Fig. 3. The measured spectrum for (a) 180° phase shift, (b) 90° phase shift, and (c) 0° phase shift.

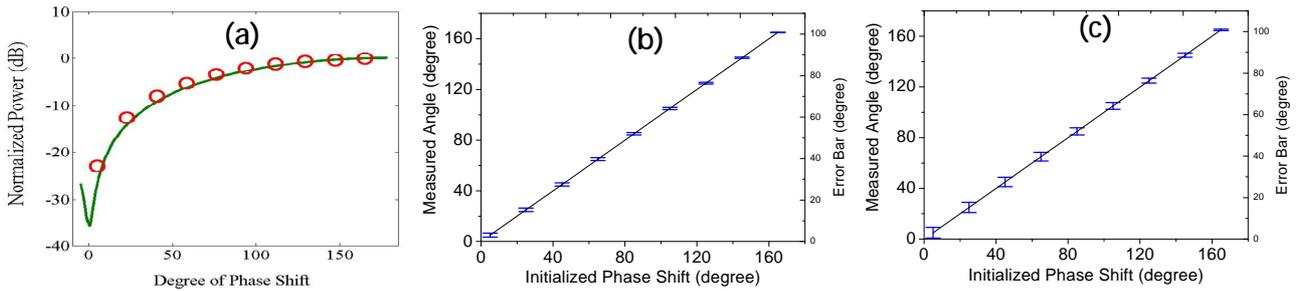

Fig. 4. (a) Measured optical power (circles) and theoretical trend (curve), $P_{mo}$=-45.3dBm; (b) Measured phase shift (dots) and their measurement errors (vertical bars), $P_{mo}$=-45.3dBm; (c) Measured phase shift (dots) and their measurement errors (vertical bars), $P_{mo}$=-40.5dBm.

In this letter, a phase modulation parallel optical delay detector for microwave angle-of-arrival measurement with accuracy monitoring is proposed using only one dual-electrode MZM. The spatial delay measurement is translated into the phase shift between two replicas of a microwave signal. Thanks to the accuracy monitoring, the phase shifts from 5° to 165° are measured with 3.1° measurement error. With the capability of accuracy monitoring, and robust parallel and simple structure, the proposed scheme can be an attractive solution for photonic AOA measurement.

This work is supported by The Netherlands Organization for Scientific Research (NWO) under the project grant Smart Optical-Wireless In-home Communication Infrastructure (SOWICI).

**References**
1. M. Jarrahi, T. H. Lee, and D. A. B. Miller, IEEE Photon. Technol. Lett., 20, 517-519(2008).
2. H. Huang, S. R. Nuccio, Y. Yue, J. Yang, Y. Ren, C. Wei, G. Yu, R. Dinu, D. Parekh, C. J. Chang-Hasnain, and A. E. Willner, J. Lightwave Technol., 30, 3647-3652( 2012).
3. E. Rouvalis, M. Chtioui, F. van Dijk, F. Lelarge, M. J. Fice, C. C. Renaud, G. Carpintero, and A. J. Seeds, Opt. Express, 20, 20090-20095( 2012).
4. H. Ito, S. Kodama, Y. Muramoto, T. Furuta, T. Nagatsuma, and T. Ishibashi, J. Sel. Top. Quantum. Electron., 10, 709-727( 2004).
5. R. K. Mohan, Z. W. Barber, C. Harrington, and W. R. Babbitt, OFC/NFOEC 2010, p. OWF3.
6. B. Vidal, M. Á. Piqueras, and J. Martí, J. Lightwave Technol., 24, 2741(2006).
7. S. Pan, J. Fu, and J. Yao, Opt. Lett., 37, 7-9(2012).
8. X. Zou, W. Li, W. Pan, B. Luo, L. Yan, and J. Yao, Opt. Lett., 37, 755-757( 2012).
9. Z. Cao, H. P. A. van den Boom, R. Lu, Q. Wang, E. Tangdiongga, and A. M. J. Koonen, IEEE, Photon. Technol. Lett., 25, 1932-1935(2013)

**Full References**

[1] M. Jarrahi, T. H. Lee, and D. A. B. Miller, "Wideband, Low Driving Voltage Traveling-Wave Mach Zehnder Modulator for RF Photonics", Photonics Technology Letters, IEEE, vol. 20, pp. 517-519, 2008.

[2] H. Huang, S. R. Nuccio, Y. Yue, J. Yang, Y. Ren, C. Wei, G. Yu, R. Dinu, D. Parekh, C. J. Chang-Hasnain, and A. E. Willner, "Broadband Modulation Performance of 100-GHz EO Polymer MZMs", Lightwave Technology, Journal of, vol. 30, pp. 3647-3652, 2012.

[3] E. Rouvalis, M. Chtioui, F. van Dijk, F. Lelarge, M. J. Fice, C. C. Renaud, G. Carpintero, and A. J. Seeds, "170 GHz uni-traveling carrier photodiodes for InP-based photonic integrated circuits", Opt. Express, vol. 20, pp. 20090-20095, 2012.

[4] H. Ito, S. Kodama, Y. Muramoto, T. Furuta, T. Nagatsuma, and T. Ishibashi, "High-speed and high-output InP-InGaAs unitraveling-carrier photodiodes", Selected Topics in Quantum Electronics, IEEE Journal of, vol. 10, pp. 709-727, 2004.

[5] R. K. Mohan, Z. W. Barber, C. Harrington, and W. R. Babbitt, "Frequency Resolved Angle and Time Difference of Arrival Estimation with Spatial Spectral Holography", in Proceedings of Optical Fiber communication/National Fiber Optic Engineers Conference 2010, OFC/NFOEC 2010, p. OWF3.

[6] B. Vidal, M. Á. Piqueras, and J. Martí, "Direction-of-Arrival Estimation of Broadband Microwave Signals in Phased-Array Antennas Using Photonic Techniques", Lightwave Technology, Journal of, vol. 24, p. 2741, 2006.

[7] S. Pan, J. Fu, and J. Yao, "Photonic approach to the simultaneous measurement of the frequency, amplitude, pulse width, and time of arrival of a microwave signal", Opt. Lett., vol. 37, pp. 7-9, 2012.

[8] X. Zou, W. Li, W. Pan, B. Luo, L. Yan, and J. Yao, "Photonic approach to the measurement of time-difference-of-arrival and angle-of-arrival of a microwave signal", Opt. Lett., vol. 37, pp. 755-757, 2012.

[9] Z. Cao, H. P. A. van den Boom, R. Lu, Q. Wang, E. Tangdiongga, and A. M. J. Koonen, "Angle-of-Arrival Measurement of a Microwave Signal Using Parallel Optical Delay Detector", Photonics Technology Letters, IEEE, vol. 25, pp. 1932-1935, 2013.